\documentclass[prl,twocolumn,showpacs,preprintnumbers,amsmath,amssymb,mathptmx,citeautoscript,aps,prl]{revtex4-1}    

\usepackage{graphicx}
\usepackage{dcolumn}
\usepackage{bm}
\usepackage{color}
\usepackage{ulem}
\usepackage{cancel}
\usepackage{enumerate}

\begin{document}

\title{Magnetic and superconducting properties of the heavy-fermion CeCoIn$_5$ epitaxial film probed by nuclear quadrupole resonance}

\author{Takayoshi Yamanaka$^1$}
\author{Masaaki Shimozawa$^2$}
\author{Hiroaki Shishido$^3$}
\author{Shunsaku Kitagawa$^1$}
\author{Hiroaki Ikeda$^4$}
\author{Takasada Shibauchi$^5$}
\author{Takahito Terashima$^1$}
\author{Yuji Matsuda$^1$}
\author{Kenji Ishida$^1$}

\affiliation{
$^1$ Department of Physics, Kyoto University, Kyoto 606-8502, Japan \\
$^2$ Institute for Solid State Physics, the University of Tokyo, Kashiwa 277-8581, Japan \\
$^3$ Department of Physics and Electronics, The School of Engineering, Osaka Prefecture University, Sakai 599-8531, Japan\\
$^4$ Department of Physics, Ritsumeikan University, Kusatsu 525-8577, Japan\\
$^5$ Department of Advanced Materials Science, Graduate School of Frontier Sciences, The University of Tokyo, Kashiwa 277-8561, Japan
}

\date{\today}

\begin{abstract}
Since the progress in the fabrication techniques of thin-films of exotic materials such as strongly correlated heavy-fermion compounds, microscopic studies of the magnetic and electronic properties inside the films have been needed. Herein, we report the first observation of $^{115}$In nuclear quadrupole resonance (NQR) in an epitaxial film of the heavy-fermion superconductor CeCoIn$_5$, for which the microscopic field gradient within the unit cell as well as magnetic and superconducting properties at zero field are evaluated. We find that the nuclear spin-lattice relaxation rate in the film is in excellent agreement with that of bulk crystals, whereas the NQR spectra show noticeable shifts and significant broadening indicating a change in the electric-field distribution inside the film. The analysis implies a displacement of In layers in the film, which however does not affect the magnetic fluctuations and superconducting pairing. This implies that inhomogeneity of the electronic field gradient in the film sample causes no pair breaking effect.
\end{abstract}

\maketitle

Recently developed thin-film fabrication techniques, such as molecular beam epitaxy (MBE) and pulsed laser deposition, can be used to fabricate films consisting of not only simple substances but also multi-element materials. 
In thin-films, it is expected that the electronic states can be manipulated by effects that are difficult to achieve in bulk systems, such as two-dimensionality \cite{Dolan1979PRL, Kobayashi1980JPSJ, McGinnis1981PhysicaBC, Burns1984SSC}, surface states \cite{Nicolay2001PRB, Sakamoto2010PRB}, and proximity effects between the sample and the substrate \cite{Anwar2015APE}. 
Recently, epitaxial thin film and superlattice samples of heavy-fermion (HF) Ce$M$In$_5$ ($M = $ Co, Rh) have been fabricated. 
The superlattices consisting of Ce$M$In$_5$ / normal-metal Yb$M$In$_5$ have attracted interest for their anomalous features such as enhancement of the anisotropy of the superconducting (SC) upper critical field ($H_{{\mathrm c}2}$) \cite{Mizukami2011NatPhys, Goh2012PRL, Shimozawa2014PRL}, suppression of antiferromagnetic (AFM) fluctuations at the interfaces between block-layers in CeCoIn$_5$/YbCoIn$_5$ superlattices \cite{Yamanaka2015PRB}, and tuning of quantum criticality by dimensionality in CeIn$_3$/LaIn$_3$ \cite{Shishido2010Science} and CeRhIn$_5$/YbRhIn$_5$ \cite{IshiiPRL2016} superlattices. 

Although such attractive phenomena have been reported, unfortunately the experimental techniques to study the physical properties of film samples are quite limited. 
For example, neutron scattering experiments, which clarify magnetic properties decisively, usually require a huge volume of the sample, typically around the order of 1 cm$^3$. 
Specific-heat and thermal-conductivity measurements, which are powerful tools to determine the SC gap structure, are also difficult since the huge contribution from the substrate masks the small contribution arising from the film sample. 

Nuclear quadrupole resonance (NQR) and nuclear magnetic resonance (NMR) are quite suitable measurements for film samples, if the material includes the NMR/NQR possible nuclei, because the measurements can detect the electronic state in only the film. 
In NMR/NQR experiments, the small sample volume reduces the signal intensity but the large surface area partially compensates the loss of signal intensity because NMR/NQR measurements mainly detect the surface region with several microns of depth.
In addition, the experiments are able to select interesting block layers of the superlattice samples and the atomic sites in the crystal structure, and thus are free from the substrate if appropriate atoms are chosen.
Particularly an NQR measurement is possible in the absence of magnetic fields and gives us the magnetic and electronic information in the SC state without the disturbance of the magnetic field.

Bulk CeCoIn$_5$ is a well-known HF superconductor with a critical temperature $T_{\mathrm c} = 2.3$ K, which is the highest $T_{\mathrm c}$ among Ce-based HF superconductors \cite{Petrovic2001JPCM}.
Many experiments have suggested that the SC gap symmetry is $d_{x^2-y^2}$-wave \cite{Izawa2001PRL, An2010PRL, Allan2013NatPhys} and a few percents of the nonmagnetic Cd- and Hg-substitutions \cite{Nicklas2007PRB, Booth2009PRB} for the In site induce antiferromagnetic (AFM) ordering with suppression of the superconductivity, indicating that CeCoIn$_5$ is located near the AFM quantum critical point. 
However, the nature of pair-breaking effect of chemical substitutions in CeCoIn$_5$ remains controversial \cite{Gofryk2012PRL, Sakai2015PRB}. 

The motivation of the present measurements is to check whether the NQR measurement is possible for the epitaxial film sample, and to investigate the magnetic and SC properties of the epitaxial film sample microscopically if NQR signals can be observed.
In epitaxial films, the strain due to the mismatch between the substrate and the films in general leads to some inhomogeneity in the electric field distribution, and thus the NQR studies will provide important information on the electronic properties and the pair-breaking effect of such inhomogeneity. 
We succeeded in observing $^{115}$In-NQR signals on the epitaxial film sample and measured the nuclear spin-lattice relaxation rate ($1/T_1$) down to 100 mK. 
We found that $1/T_1$ in the normal and SC states is very similar between the bulk and film samples, although the $^{115}$In-NQR spectrum on the film sample shifts noticeably and becomes significantly broader than the bulk sample.
These results indicate that the magnetic and superconducting properties, including the gap structure of the epitaxial film sample, are identical to the bulk sample although some modulation of the crystal structure is introduced; thus, the NQR measurement is a useful technique to investigate the electronic state and SC gap in film samples, in which the specific heat and thermal conductivity measurements are almost impossible.
As far as we know, these are the first NQR measurements of film samples of strongly correlated electron systems. 

The CeCoIn$_5$ epitaxial film studied here was grown along the tetragonal $c$-axis using MBE on a MgF$_2$ substrate (See Ref.\cite{Mizukami2011NatPhys, Shimozawa2012PRB} for fabrication details). 
The electrical resistivity $\rho (T)$ of a 120-nm-thick film of CeCoIn$_5$ was measured, and exhibited almost the same behavior as bulk single-crystal CeCoIn$_5$; the maximum $\rho (T)$ associated with an incoherent-coherent crossover of $f$-electrons was observed at around 40 K and the $T_{\mathrm{c}} = 1.95\; \mathrm{K}$ of the film sample is close to the bulk $T_{\mathrm{c}}$ \cite{Shimozawa2012PRB}.
The epitaxial film prepared for the NQR measurements was 500 nm thick, which approximately corresponds to 600 unit-cell thickness, and the surface area was  8 $\times$ 5 mm$^2$. 
$T_{\mathrm{c}} = 2.15\; \mathrm{K}$ was determined with the NQR 1/$T_1$ measurement.  
For comparison, we have performed $^{115}$In-NQR on bulk single-crystal CeCoIn$_5$ without powdering, which was grown with the In self-flux method.
The size of the crystal was $4 \times 3 \times 0.5 \mathrm{mm}^3$.
For NQR measurements below 1.5 K, a $^3$He-$^4$He dilution refrigerator was used and the samples were immersed into the $^3$He-$^4$He mixture to avoid any heating by RF pulses for the NQR measurements.
$1/T_1$ was measured from the recovery of the nuclear magnetization $m(t)$ at a time $t$ after a saturation pulse. 

\begin{figure}
\includegraphics[width=1.0\linewidth]{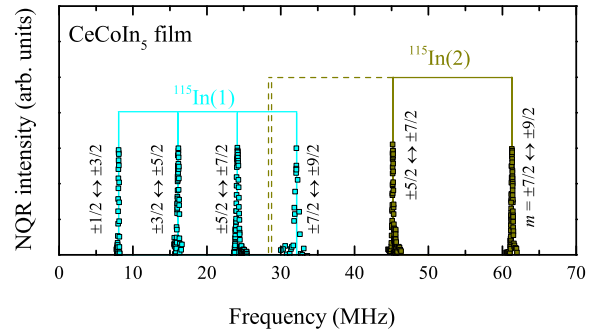}
\caption{NQR spectra arising from $^{115}$In(1) and $^{115}$In(2) of the film CeCoIn$_5$ at $T \sim 2.5$ K.
The intensity of each signal is normalized by each maximum. 
The dashed lines indicate the frequencies where observation of NQR signals are expected from the NQR parameters. 
} 
\label{fig_SP}
\end{figure}

\begin{figure}
\includegraphics[width=0.98\linewidth]{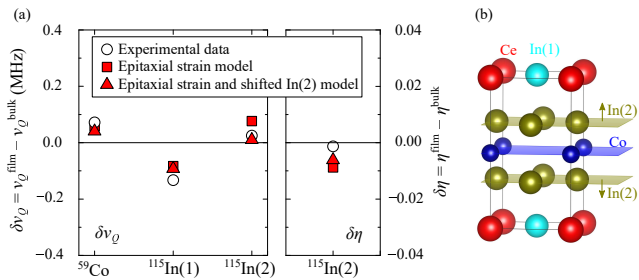}
\caption{(a)Differences of the $\nu _Q$ between the bulk and film CeCoIn$_5$(see the main text). 
Open circles show experimental data, red squares show the epitaxial strain effect, and the red triangles show the epitaxial strain and In(2) shift effects. 
As for the In(2) shift effect, the Wyckoff parameter of In(2) is changed from 0.3094 to 0.3090.
The EFG parameters of the red symbols were calculated with WIEN2k \cite{SMRef}.  (b) Image of the In(2) shift effect.} 
\label{fig_calc_nuQ}
\end{figure}

Figure \ref{fig_SP} shows the $^{115}$In NQR spectrum measured in the CeCoIn$_5$ film. 
CeCoIn$_5$ has two In sites denoted as In(1) and In(2); In(1) is located at the center of the Ce-In plane and In(2) is located between the Co and the Ce-In(1) planes on the lateral plane(Fig. \ref{fig_calc_nuQ}(b)). 
Four lines at 8.05 MHz, 16.11 MHz, 24.16 MHz and 32.17 MHz arise from the In(1) site and the other two lines at 45.173 MHz and 61.310 MHz arise from the In(2) site. 
The electric quadrupole Hamiltonian is written as 
\[ \mathcal{H} _Q = \frac{e^2 qQ}{4I(2I-1)} \left[ 3\hat{I}_z^2-\hat{I}^2 + \frac{\eta}{2} \left( \hat{I}_+ ^2 + \hat{I}_- ^2 \right) \right], \]
where $Q$ is the nuclear quadrupole moment. 
The electric field gradient (EFG) originating from the electrons and atoms surrounding the $^{115}$In nucleus is denoted as $eq \equiv V_{zz}$, and the asymmetric parameter $\eta$ is defined as $\eta \equiv (V_{xx}-V_{yy})/V_{zz}$, where $V$ is the electric potential at the $^{115}$In site and $V_{\alpha \alpha} \equiv \frac{\partial ^2 V}{\partial \alpha ^2} (\alpha = x, y, \mathrm{and}\;z)$.
The $z$-axis is the principal axis of the EFG, $i,e.$ the $V_{zz}$ is maximum; $x$- and $y$-axes are chosen so that $0 \leq \eta \leq 1$.
For the $I = 9/2$ ($^{115}$In) case, four transitions are allowed.
$^{115}$In for $\eta = 0$, a set of lines would be observed at the resonant frequencies of $\nu _Q \equiv e^2 q Q/24h, 2\nu _Q , 3\nu _Q, \mathrm{and}\;4 \nu _Q$, where $h$ is the Planck constant.
At the axially symmetric In(1) site, where the $z$-axis points to the tetragonal $c$ axis, the observed four lines were reproduced by the calculation with $\nu _Q = 8.05$ MHz and $\eta < 0.003$. 
On the other hand, from the set of lines arising from the asymmetric In(2) site, where the $z$-axis points perpendicularly to the lateral plane of the unit cell, $\nu _Q = 15.52$ MHz and $\eta = 0.39$ were evaluated.
These NQR parameters $\nu_Q$ and $\eta$ of the epitaxial film sample are listed in Table \ref{tab_EFG} and are approximately the same as those of bulk samples \cite{Kohori2001PRB, Curro2001PRB}.

\begin{figure}
	\begin{minipage}{0.48\linewidth}
		\includegraphics[width=0.95\linewidth]{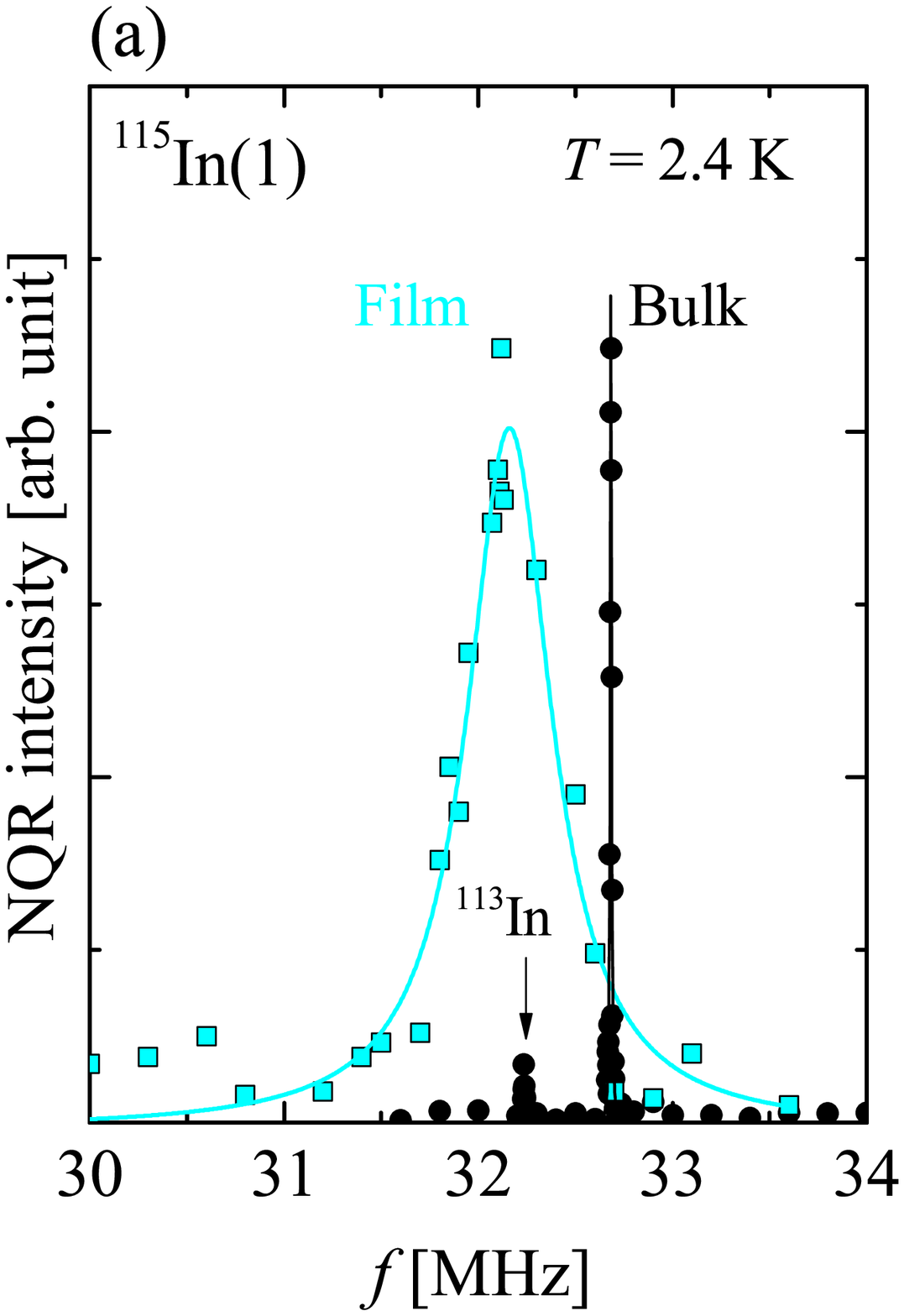}
	\end{minipage}
	\begin{minipage}{0.48\linewidth}
		\includegraphics[width=0.95\linewidth]{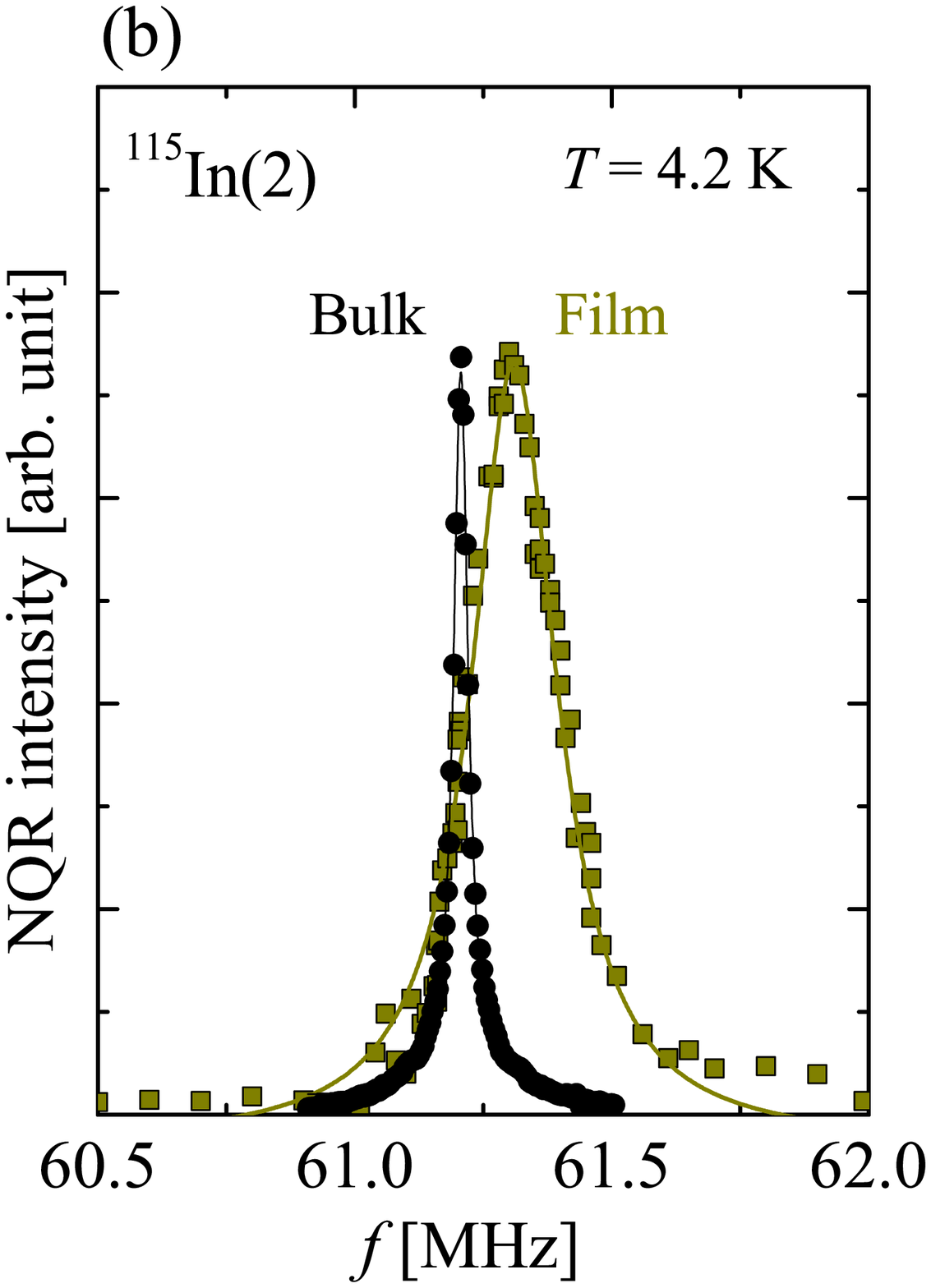}
	\end{minipage}
\caption{NQR spectra for transition between $m = \pm 7/2\;\mathrm{and}\;\pm 9/2$\; of the In(1) and the In(2) in the film CeCoIn$_5$ (squares) and bulk single-crystal (circles). 
The solid lines are fittings of the spectrum to the Lorentzian function. 
The fitting parameters ($f_\mathrm{center}$, FWHM) were evaluated as (32.16 MHz, 530 kHz) for In(1), and (61.31 MHz, 203 kHz) for In(2) in the film sample.
In the bulk sample, ($f_\mathrm{center}$, FWHM) were (32.68 MHz, 8 kHz) for In(1), and (61.21 MHz, 35 kHz) for In(2).  
The intensity of the NQR signal of the film was magnified 60 times.
In the bulk single-crystal spectrum of In(1), the tiny signal at 32.24 MHz close to the film NQR frequency is ascribed to the isotope $^{113}$In \cite{SMRef}}. 

\label{fig_SP_4nuQ}
\end{figure}

Figure \ref{fig_SP_4nuQ} is an expanded view around the signals of the transition of $m = \pm 7/2 \leftrightarrow \pm 9/2$ at the $^{115}$In(1) and $^{115}$In(2) sites in the film and single-crystal CeCoIn$_5$.
The peaks of the film NQR spectrum were clearly shifted from those of the single-crystal spectrum.
In addition to the noticeable frequency shift, the line width of the film sample is much broader than that of the single-crystal sample. 
The origin of the spectrum shift and broadening of the NQR spectrum are discussed later.  

\begin{table}
\caption{EFG parameters of the film and single crystalline samples in the present measurement. The single crystalline CeCoIn$_5$ results reported by Curro {\it et al.} \cite{Curro2001PRB} are also shown. The $^{59}$Co data were measured by NMR experiments.}
\begin{tabular}{c|cccc}
\hline \hline 
 & $^{59}$Co & $^{115}$In(1)  & \multicolumn{2}{c}{$^{115}$In(2)}  \\
Sample & $\nu _Q$ (MHz) & $\nu _Q$ (MHz) & $\nu _Q$ (MHz) & $\eta $ \\ \hline
Film & 0.302 \cite{Yamanaka2015PRB} & 8.05 & 15.52 & 0.39 \\
Bulk & 0.230 & 8.171 & 15.491& 0.387\\ 
Bulk \cite{Curro2001PRB} & 0.234 & 8.173 & 15.489 & 0.386 \\ \hline \hline
\end{tabular}

\label{tab_EFG}

\end{table}

\begin{figure}

\includegraphics[width=0.75\linewidth]{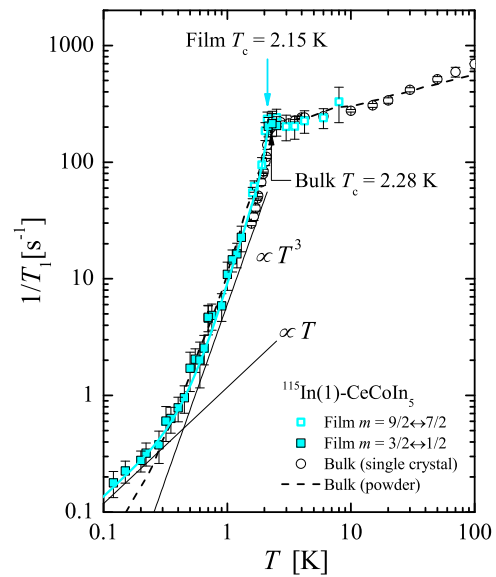}

\caption{Temperature dependence of $1/T_1$ at the In(1) site measured at the transition of $m = \pm 7/2 \leftrightarrow \pm 9/2$ (open squares) and $m = \pm 1/2 \leftrightarrow \pm 3/2$ (solid squares) of the film sample. The blue solid line is the calculation by the $d_{x^2-y^2}$-wave SC gap model with the parameters of $2\Delta(0)/k_\mathrm{B}T_\mathrm{c}=6$ and $N_\mathrm{Res}/N(T_0)=0.12$.
The open circles are the data of transition of $m = \pm 7/2 \leftrightarrow \pm 9/2$ in the bulk single crystal. The dashed line represents the data set of powdered CeCoIn$_5$ reported by Kohori, {\it et al}. \cite{Kohori2001PRB}. 
}
\label{fig_T1}

\end{figure}

We measured $1/T_1$ at the In(1) site in order to investigate magnetic and SC properties of the film sample.
$1/T_1$ in the film sample could be determined with a single component in the whole temperature range, indicating that magnetic fluctuations are homogeneous in the whole region of the sample.   
The temperature dependence of $1/T_1$ in the film (squares) and the bulk single-crystal (circles) is compared with the data of crushed single-crystal (dashed line) CeCoIn$_5$ reported by Kohori {\it et al}. \cite{Kohori2001PRB} in Fig. \ref{fig_T1}.
The $T$ dependence of $1/T_1$ of the present film sample shows the same behavior as the bulk single crystals reported previously \cite{Kohori2001PRB, Kawasaki2003JPSJ}.
$1/T_1$ in the normal state deviates from the Korringa relation observed in conventional metals, indicative of the presence of the antiferromagnetic correlations. 
Below $T_c$, $1/T_1$ decreases rapidly without a coherence peak just below $T_\mathrm{c} = 2.15$ K, which is close to the bulk $T_\mathrm{c} = 2.28$ K, and  $1/T_1$ for $0.4\;\mathrm{K} < T < 1.5$ K obeys $T^3$-like behavior expected in the line-node SC gap model. 
In addition, $1/T_1$ below $0.4\;\mathrm{K}$ deviates from the $T^3$ dependence and approaches a $T$-linear behavior. 
Such a $T$-linear behavior at the lowest temperatures has been observed in various unconventional superconductors with a sign-changed line-node gap and implies the presence of a residual density of states (RDOS) $N_\mathrm{Res}$ at the Fermi level \cite{Monien1990PRB,Ishida1993JPSJ, Tokunaga2001PhysicaC}.
These results indicate that the SC gap symmetry of the 500-nm-thick epitaxial film CeCoIn$_5$ is also $d$-wave, similarly to bulk single-crystal CeCoIn$_5$.

The temperature dependence of $1/T_1$ below $T_\mathrm{c}$ is consistently understood with the 2-dimensional $d_{x^2-y^2}$-wave SC gap model with $\Delta(T,\phi) = \Delta(T)\cos (2\phi)$, in which the magnitude of the SC gap $2\Delta(0)/k_\mathrm{B}T_\mathrm{c}$ and the fraction of RDOS  $N_\mathrm{Res}/N_0$ are variable parameters\cite{Hotta1993JPSJ}. 
Here $N_0$ is the density of states in the normal state.
The ratio of $1/T_{1}$ in the SC state to $1/T_{1}$ at $T=T_{\mathrm c}$ is expressed as
\begin{align*}
\frac{(1/{T_{1}})}{(1/T_1)_{T=T{\mathrm c}}}=\frac{2}{k_\mathrm{B}T_\mathrm{c}}\int ^\infty _0 \frac{N_\mathrm{s}(E)^2 +M^2 }{N_0^2} \left(-\frac{\partial f(E)}{\partial E}\right) dE,
\end{align*}
where $f(E)$ is the Fermi distribution function, $N_\mathrm{s}(E)$ is the density of states in the SC state,
and $M$ is the anomalous density of states, which vanishes in the sign-changed SC gap. 
The data of $1/T_{1}$ is well reproduced by the calculation with the parameters of $2\Delta(0)/k_\mathrm{B}T_\mathrm{c}=6$ and $N_\mathrm{Res}/N_0 =0.12$ as shown in Fig. \ref{fig_T1}\cite{SMRef}.
Although the magnitude of the SC gap in the film sample is the same as that in the single-crystal sample, the fraction of RDOS in the film sample is slightly larger than that in the single-crystal sample ($N_\mathrm{Res}/N_0 \sim 0.08$) \cite{Kohori2001PRB,Yashima2004JPSJ}. 
In general, randomness or imperfection, which is a perturbation averaging the $k$-dependence of the SC gap, induces the RDOS near the nodes. 
The larger RDOS in the film sample than in the bulk single-crystal sample is consistent with the broader NQR spectrum in the former sample.
However, this effect is much weaker than the effect by the nonmagnetic-impurity doping, because the reduction of $T_c$ in the film sample is only 6\%.     

Now, we discuss the change of the NQR spectra in the epitaxial film sample.
As shown above, we observed a noticeable shift and significant broadening of the $^{115}$In-NQR spectra in the film sample. 
The changes of the NQR frequency $\nu_Q$ from the bulk single crystal $\delta \nu _Q (= \nu _Q^\mathrm{film}-\nu _Q^\mathrm{bulk})$ are plotted in Fig. \ref{fig_calc_nuQ}(a), where the change of the $\nu_Q$ at the Co site and the change of the $\eta$, $\delta \eta (= \eta ^\mathrm{film}-\eta ^\mathrm{bulk})$ at the In(2) site are also plotted. 
In addition, the full width at half maximum (FWHM) of the spectra for the film sample shown in Fig. \ref{fig_SP_4nuQ} (a) and (b) are 530 kHz and 203 kHz, which are 60 times and 6 times larger, respectively, than in the bulk single crystal sample.

We now consider the origin of the difference of NQR spectra between the film and single-crystal CeCoIn$_5$.
First, there is a gradual epitaxial strain introduced by the misfit of the lattice parameters between CeCoIn$_5$ and the MgF$_2$ substrate.  
The lattice parameters of the 120-nm-thick epitaxial film sample were reported to be $a = 0.462$ nm and $c = 0.753$ nm \cite{Shimozawa2012PRB}, which were slightly different from those of the bulk single crystal ($a = 0.461$ nm and $c = 0.755$ nm). 
Using these film lattice parameter, the EFG parameters by the band calculation \cite{SMRef} approximately reproduce the $\delta \nu _Q$ and $\delta \eta$ (squares in Fig. \ref{fig_calc_nuQ}(a)). 
In addition, we introduce the shift of the In(2) layers toward the Ce-In(1) layers as shown in Fig. \ref{fig_calc_nuQ}(b), and the EFG parameters including the two effects are more consistent with the experimental data (triangles in Fig. \ref{fig_calc_nuQ}(a)). 
In general, an epitaxial strain is an inevitable effect near a substrate and is relaxed with the distance away from the substrate, and the lattice parameters far from the substrate are considered to be the same as those in the bulk sample. 
Such a gradation of the lattice parameters would cause an asymmetric distribution of the $\nu _Q$ from a bulk NQR peak, but the symmetric spectrum was observed in the film CeCoIn$_5$. 
Alternatively, we suggest that the lattice parameters in epitaxial film samples might be optimized by the condition in which the strain effect and other effects are included.


Although significant broadening of the NQR signals is introduced in the film sample, it is surprising that the $T_\mathrm{c}$, $1/T_1$, and the RDOS of the film sample are essentially the same as those in the bulk single-crystal CeCoIn$_5$. 
The present results are consistent with the experimental fact that the resistivity and Hall effect in the 120-nm-thick film CeCoIn$_5$ are nearly identical to  those in the single crystal \cite{Shimozawa2012PRB}, but they seem to be contradict with the fact that the superconductivity in CeCoIn$_5$ is easily disrupted by the nonmagnetic impurities\cite{Nicklas2007PRB, Booth2009PRB, Bauer2006PRB}, because CeCoIn$_5$ is a quantum critical superconductor.
In the substitution systems of Cd (hole doping) [CeCo(In$_{1-x}$Cd$_x$)$_5$], the superconductivity is destroyed at the critical concentration of $x = 1.7$\% and long-range AFM ordering appears.
1\% Cd-doping induces the small satellite structure in the $^{115}$In-NQR spectrum \cite{Urbano2007PRL, Seo2014NatPhys, Sakai2015PRB}, and largely modifies the spin-fluctuation properties in the normal state near the impurity sites.

In Sn-doing (electron doping), $\nu _Q$ of the main peak at the In(1) site decreases similarly to our film sample but the $1/T_1$ drastically decreases by enhanced $p$-$f$ hybridization due to the additional $5p$ electron of Sn atoms \cite{Sakai2015PRB}.
In these doping, impurities on CeCoIn$_5$ not only scatter the conduction electrons, but also modify their $c$-$f$ hybridization\cite{Gofryk2012PRL,Sakai2015PRB}.
In the film case, the primary difference from the bulk sample is the disorder in the electric field gradient, and our results showing identical $1/T_1$ with the bulk one indicate that the band structure around the Fermi level remains almost unchanged from the bulk band structure. 
From the comparison between the film sample and the nonmagnetic impurity-doped samples, we suggest that the $\nu _Q$ distribution effect, which is introduced by the inhomogeneity of the tiny lattice change and In(2) shift, does not seriously alter the magnetic and SC properties.


In conclusion, we performed NQR experiments in 500-nm-thick epitaxial film CeCoIn$_5$ and in bulk single crystal, and found the following two points.
(a) Although the $\nu _Q$ distribution is introduced in the epitaxial film sample, $T_\mathrm{c}$, $1/T_1$, and the RDOS of the film sample are essentially the same as those in the single-crystal CeCoIn$_5$, indicating that the magnetic and SC properties in the film sample are identical to the single crystal sample. Thus, an NQR measurement is a suitable technique to investigate the SC gap in film samples.
(b) The origin of the spectrum broadening is partially interpreted by the epitaxial strain effect, but we suggest that the In(2) layer slightly shifts to the Ce-In(1) layer in the epitaxial film sample. We suggest from the present NQR study that the lattice parameters might be optimized in the epitaxial film sample.

\noindent
{\bf Acknowledgments} 
The authors acknowledge M. Yashima for fruitful discussions. 
This work was partially supported by Kyoto Univ. LTM center, Grant-in-Aid from the Ministry of Education, Culture, Sports, Science, Technology(MEXT) of Japan, Grants-in-Aid for Scientific Research (KAKENHI) from Japan Society for the Promotion of Science (JSPS), ``J-Physics'' (No.\,JP15H05882, No.\,JP15H05884, and No.\,JP15K21732) and ``Topological Quantum Phenomena'' (No.\,JP25103713) Grant-in-Aid for Scientific Research on Innovative Areas from the MEXT of Japan, and by Grant-in-Aids for Scientific Research (Grants No. JP25220710, and No. JP15H05745).
Figures of crystal structures were shown using the VESTA package \cite{vesta}.

%

\end{document}


\title{Supplemental Material; Magnetic and superconducting properties of the heavy-fermion thin film probed by nuclear quadrupole resonance}

\author{Takayoshi Yamanaka$^1$}
\author{Masaaki Shimozawa$^2$}
\author{Hiroaki Shishido$^3$}
\author{Shunsaku Kitagawa$^1$}
\author{Hiroaki Ikeda$^4$}
\author{Takasada Shibauchi$^5$}
\author{Takahito Terashima$^1$}
\author{Yuji Matsuda$^1$}
\author{Kenji Ishida$^1$}

\affiliation{
$^1$ Department of Physics, Kyoto University, Kyoto 606-8502, Japan \\
$^2$ Institute for Solid State Physics, the University of Tokyo, Kashiwa 277-8581, Japan \\
$^3$ Department of Physics and Electronics, The School of Engineering, Osaka Prefecture University, Sakai 599-8531, Japan\\
$^4$ Department of Physics, Ritsumeikan University, Kusatsu 525-8577, Japan\\
$^5$ Department of Advanced Materials Science, Graduate School of Frontier Sciences, The University of Tokyo, Kashiwa 277-8561, Japan
}

\date{\today}
\maketitle
\section{NQR signals of  Indium isotopes}

Here, we discuss the tiny signal in Fig. 2(a) in the main text and show the evidences that the tiny signal arises from $^{113}$In isotope.
In the case of $I=9/2$, and $\eta = 0$ such as the symmetric In(1) site (Fig. 4 (b) in the main text), the resonant frequency for the transition between $m = \pm 7/2 \leftrightarrow \pm 9/2$ is written as $4\nu _Q =e^2qQ/6h$ and therefore is proportional to nuclear electric quadrupole moment $Q$ under a certain electric field gradient $eq$.
In the NQR spectrum of bulk CeCoIn$_5$, the signals were observed at 32.24 MHz(the tiny signal) and 32.68 MHz(the large signal).
The ratio of resonant frequency $32.24/32.68 \approx 0.987$ is almost the same as that of the quadrupole moment of $^{113}$In and $^{115}$In; $Q_{113}/Q_{115}=0.759/0.770 \approx 0.986$\cite{EQMTable}.

The intensity of the tiny signal is about 18 times smaller than the large ($^{115}$In) one.
The weak intensity can be explained from the point of view of $^{113}$In signal.
The natural abundance (NA) of $^{113}$In is 4.3 \% although that of $^{115}$In is 95.7 \% \cite{NA_list}.
The ratio of the experimental NQR signal intensity is consistent with the ratio of the NA of Indium isotopes.

\begin{figure}[htb]
\includegraphics[width=0.7\linewidth ]{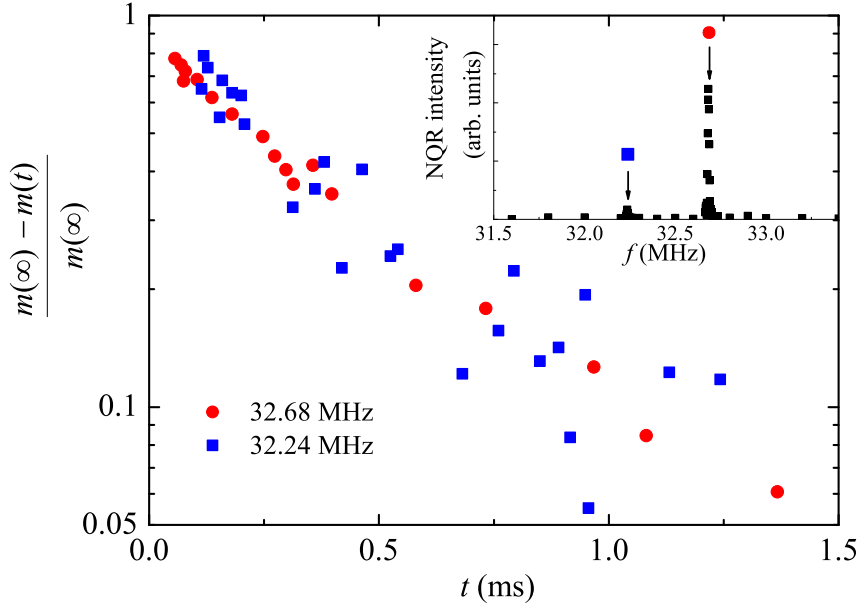}
\caption{Recovery curves of $1/T_1$ measurements. (Inset) The NQR spectraum in the normal state. The relaxation curves in the main panel were measured at each peaks pointed by arrows and symbols.}
\label{recov}
\end{figure}

The recovery curves of NQR-$1/T_1$ measurements are also consistent.
Since both of $^{113}$In and $^{115}$In have nuclear spin $9/2$, the nuclear-spin lattice relaxation follows the identical recovery curve.
In fact, both behavior of nuclear magnetization $m(t)$ at tiny and large signals are the same as shown in Fig. S \ref{recov}.
The fact suggests that the two signals arise from Indium nuclei at the same crystallographic site.

\section{NQR-$1/T_1$ in superconducting state analysis}
In almost metallic samples, conduction electron spins strongly contribute to nuclear spin relaxation and therefore nuclear-spin lattice relaxation rate ($1/T_1$) depends on density of states (DOS) at the Fermi level.
In the SC state, the nuclear spins are relaxed by thermally excited quasi particles and therefore the $1/T_1$ depends on the coherence factor.
$1/T_1$ in the SC states is written as \cite{Hebel1959PR,MacLaughlin19761,Monien1990PRB}
\begin{align*}
\frac{1/{T_{1}}}{(1/T_1)_{T=T{\mathrm c}}}&=\frac{2}{k_\mathrm{B}T_\mathrm{c}}\int ^\infty _\Delta  \frac{N_\mathrm{s}(E)^2 +M(E)^2 }{N_0^2} \left(-\frac{\partial f(E)}{\partial E}\right) dE, \\
 N_{\mathrm s}(E)/N_0 & \equiv {\Big \langle} \frac{E}{\sqrt{E^2 - \Delta ^2 (T, \theta, \phi)}}{\Big \rangle}_\mathrm{F.S.}, \\
 M_{\mathrm s}(E)/N_0 & \equiv {\Big \langle} \frac{\Delta}{\sqrt{E^2 - \Delta ^2 (T, \theta, \phi)}}{\Big \rangle}_\mathrm{F.S.}.
\end{align*}
Here, $ \langle \cdots \rangle _\mathrm{F.S.}$ is an average over the Fermi surface. 
$N_0$, and $N_\mathrm{s}(E)$ are DOS at Fermi surface in the normal and SC states, respectively, and $E$ is the energy from the Fermi level.
The $M_{\mathrm s}$ is called as anomalous DOS arising from the coherence factor, which is strictly zero in the $d$-wave symmetry because of the sign change of the SC energy gap $\Delta$.
We assumed the two dimensional $d_{x^2-y^2}$-wave gap symmetry described as $\Delta (T, \theta , \phi) = \Delta (T) \cos (2\phi)$ and then obtain
\begin{align*}
 N_{\mathrm s}(E)/N_0 &\equiv \frac{1}{4\pi}\int _0 ^\pi \sin \theta d \theta \int _0 ^{2\pi} d \phi \frac{E}{\sqrt{E^2 - \Delta ^2 (T)\cos^2 (2\phi)}}  \\
 	&=\frac{1}{2\pi}\int _0 ^{2\pi} d \phi \frac{1}{\sqrt{1 - \left[\Delta (T)/E \cos (2\phi)\right]^2}} \\
 	&=\frac{2}{\pi}\int _0 ^{\pi /2} d \phi \frac{1}{\sqrt{1 - \left[\Delta (T)/E \cos (2\phi)\right]^2 }}.
\end{align*}
At the last equation, we used $\pi /2$-periodicity of the integrand. 
In the $\Delta (T)/E \geq 1$ case, a constant $\alpha $ satisfying $\cos (2\alpha ) =E/\Delta (T)$ exists and therefore the integration is divergent.
To avoid the divergence, we numerically calculated the $N_{\mathrm s}$ under the condition of $E > \Delta (T) \cos (2\phi)$.
For the temperature dependence of $\Delta (T)$, we used the empirical representation in BCS model as follow \cite{Tou2005JPSJ}
\begin{align*}
	\Delta (T) = \Delta (0) \tanh \left[ \frac{\pi k_\mathrm{B} T_\mathrm{c}}{\Delta (0)} \sqrt{2\frac{\Delta C}{C}\left( \frac{T_\mathrm{c}}{T}-1 \right) }\right] .
\end{align*}
Here $\Delta C/C$ is the specific heat jump. Since the specific heat measurement for film sample cannot be performed, we chose $ \Delta C/C =4.5$, which is obtained in the bulk CeCoIn$_5$ \cite{Petrovic2001JPCM}.

\begin{figure}[htb]
\includegraphics[width=0.7\linewidth ]{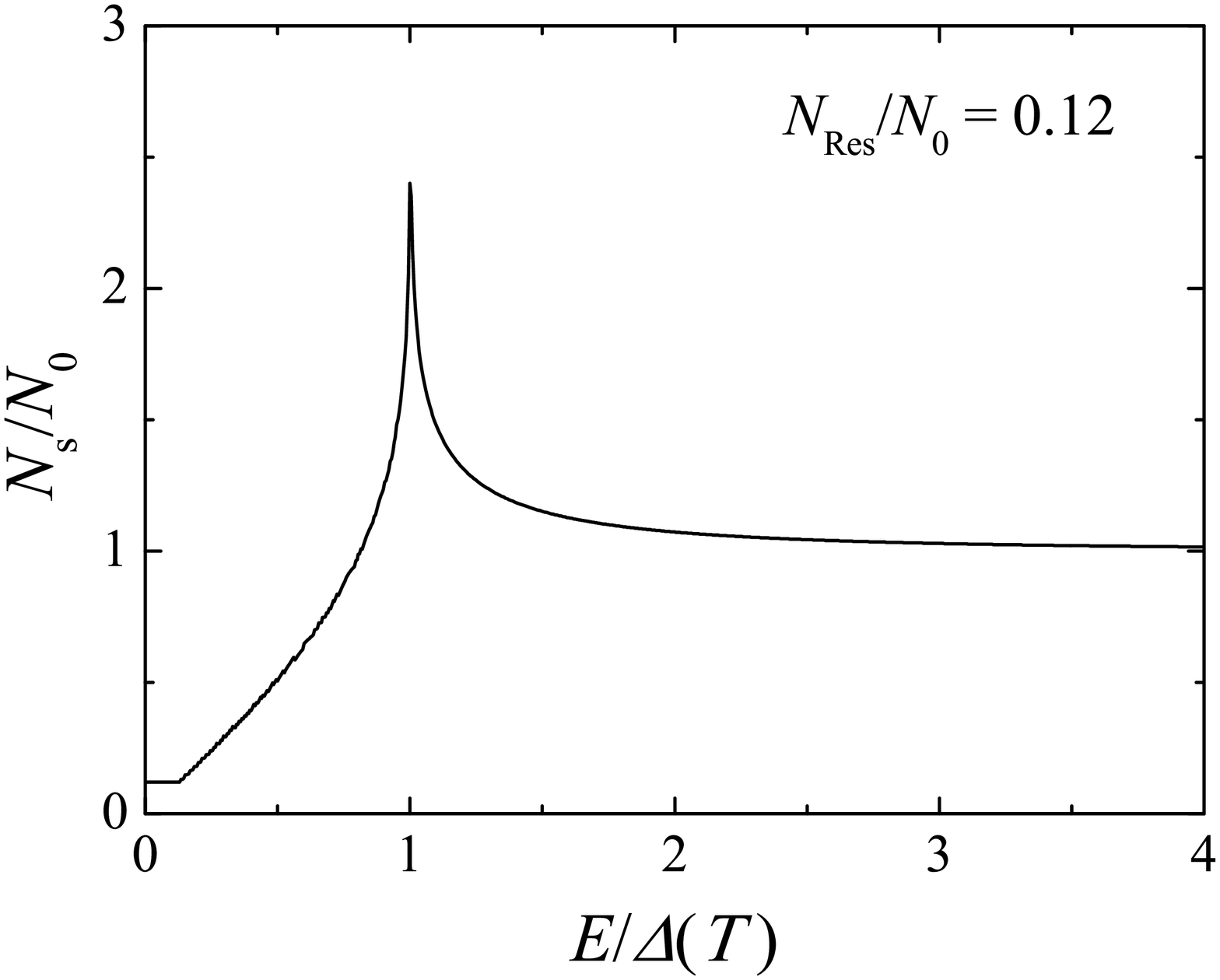}
\caption{Energy dependence of $N_\mathrm{s}$ in $d_{x^2-y^2}$-wave gap model normalized by DOS of the normal state $N_0$. 
We introduce the residual DOS $N_\mathrm{res}=0.12N_0$ to the lowest energy side.} 
\label{S_DOS}
\end{figure}

In an ideal model, the $N_{\mathrm s}(E)$ goes to zero for $E \rightarrow 0$, but in practical model of nodal SC gap, finite DOS at $E=0$ arises from impurity scattering in a low-lying energy region\cite{Schmitt-Rink1986PRL, Hotta1993JPSJ}.
We assumed that the residual DOS were constant for simplicity and calculated $N_\mathrm{s}$ including the residual DOS as shown in Fig. S \ref{S_DOS}.

\section{Electronic band structure}
The NQR parameters $\nu_Q$ and $\eta$ of the film obtained from our experiments are approximately the same as those of bulk but slightly different. 
The broadening of the NQR spectra may be due to distribution of local environment of In sites. 
In order to obtain information on the NQR spectra of a film CeCoIn$_5$, we calculated the electronic structure using the {\tt WIEN2k} package~\cite{wien2k}. 
The fully-relativistic full-potential calculations were performed with the GGA-PBE functional~\cite{PBE}, $12 \times 12 \times 7$ $k$-point grid in the Brillouin zone, and the cut-off parameter $RK_{max}=7$. 
Instead, as the first step, we examined the electronic structure of such virtual bulk CeCoIn$_5$ with the film lattice parameters, not superstructure, since the NQR parameters are sensitive to the local environment of In sites. 
The lattice parameters used here are listed in Table S\;I. 
In these virtual bulk models, the band structure is quite similar to that of the original bulk, since the modification of crystallographic parameters is small.\begin{table}[h]
\begin{tabular}{ccccccccc}
\hline \hline
& $a$ (\AA) & $c$ (\AA) & $z$ & \multicolumn{4}{c}{$V_{zz}$($10^{21}$V/m$^2$)}& $\eta$ \\ 
& & & & Ce & Co & In(1) & In(2)& In(2) \\ \hline
(bulk) & 4.612 & 7.551 & 0.3094 & 0.869 & -0.0312 & 10.889 & 19.588 & 0.278 \\ \hline
(film) & 4.620 & 7.530 & 0.3094 & 0.480 & -0.1145 & 10.790 & 19.680 & 0.268 \\
 & 4.620 & 7.530 & 0.3090 & 0.449 & -0.0899 & 10.779 & 19.600 & 0.272 \\ \hline \hline
\end{tabular}
\caption{Calculated electric field gradients of $V_{zz}$ at each atomic site, and $\eta $ at In(2) site under the set values $a$, $c$, and $z$. $a$ and $c$ are lattice constants, and $z$ is Wyckoff parameter of In(2).}
\end{table}
$\nu _{Q}$ is obtained by the formula $\nu _{Q} = 6\times \frac{eQ}{4I\cdot{2I-1}}|V_{zz}|$, in which quadrupole moment $Q$ and nuclear spin $I$ are depend on the nuclei.
The obtained NQR parameters in Table S\;I are deviated from the experimental observations due to the electron correlation effect. 
However, the difference $ \delta \nu _Q (\delta \eta)$ is comparable to those of the bulk.

%